\documentstyle[preprint,aps]{revtex}

\begin{document}

\title{The mean energy, strength and width of triple giant dipole resonances}

\author{E.J.V. de Passos\( ^{1} \), M.S. Hussein\( ^{1} \), L.F.Canto\( ^{2} \)
and B.V. Carlson\( ^{3} \)}

\address{\( ^{1} \)Instituto de F\'{\i}sica, Universidade de S\~{a}o Paulo, CP 66318,\\
 05389-970 S\~{a}o Paulo SP, Brazil\\
 \( ^{2} \)Instituto de F\'{\i}sica da Universidade Federal do Rio de Janeiro,
CP 68528,\\
 21945-970 Rio de Janeiro RJ, Brazil\\
 \( ^{3} \)Departamento de F\'{\i}sica, Instituto Tecnol\'{o}gico de\\
 Aeron\'{a}utica -- CTA, 12228-900 S\~{a}o Jos\'{e} dos Campos SP Brazil}

\maketitle
\begin{abstract}
We investigate the mean energy, strength and width of the triple giant dipole
resonance using sum rules. We point out that the excitation operator for the
triple giant dipole resonance must be modified from the naive choice \( D^{3} \).
With the modified excitation operator, the strength and mean energy calculated
in the Tomonaga model agree with the harmonic limit and the width is larger
than \( \sqrt{3} \) times that of the giant dipole resonance. Had we adopted
the naive choice, we would have obtained unphysical results for these quantities.
Assuming the absence of a charge exchange component in the nuclear interaction,
we derive model independent relationships between the mean energies and strengths
of multiple giant dipole resonances, which are an extension of relations derived
previously for the double giant dipole resonance. 
\end{abstract}

\vspace{1.0cm}

The study of the double giant dipole resonance in nuclei has received a considerable
amount of attention over the last 15 years \cite{Aum98}. Both the pion double
charge exchange and relativistic heavy ion Coulomb excitation reactions have
been used to probe this large amplitude collective motion in many fermion systems.
The quest for the similar double plasmon resonance in metallic clusters is underway
\cite{Iss99}. In nuclei, experimental investigations of the triple
giant resonances are underway \cite{Eml00,Fra99}. It is clearly
of importance to supply theoretical estimates of the cross-section as well as
the different decay branching ratios of these exotic collective modes. 

Even though they are not directly related to the cross sections, sum rules provide
useful constraints on the experimental data. These sum rules are the moments
of the strength distribution of appropriately chosen excitation operators. In
the case of the single and double giant dipole resonances, they are the dipole
and squared dipole operators, respectively. Microscopic\cite{Nis95} and more
schematic calculations \cite{Kur96} show that this choice reproduces the expected
qualitative properties of these collective states. 

In this paper, we will investigate the mean energy, strength and width of the
triple giant dipole resonance using sum rules \cite{Kur96,Boh96}. First, following
these references, we derive model independent expressions for the mean energy,
strength and width of the strength distribution of a given excitation operator
and we will give a physical interpretation of these quantities by introducing
the concept of a doorway state. Next, we discuss how to define the proper excitation
operator to probe multiple giant dipole states and point out that the excitation
operator which probes the triple giant dipole state has to be modified from
the naive choice of \( D^{3} \). We use the Tomonaga theory\cite{Kur96,Tom55}
to show that our scheme leads to the expected results for the triple giant dipole
resonance. The consequences of the hypothesis of absence of exchange forces
is also examined. At the end, we comment on the generalization to multiple giant
resonances of different multipolarities as, for example, the quadrupole resonances.

Given an excitation operator \( O^{(i)} \), we define its strength function
as \cite{Kur96,Boh96}\[
S^{(i)}(E)=\sum _{n\neq 0}\left| \left\langle n\left| O^{(i)}\right| 0\right\rangle \right| ^{2}\delta (E-\varepsilon _{n}),\]
 where \( \left| n\right\rangle ,\: n\neq 0 \) are the excited states of the
nucleus and the \( \varepsilon _{n} \) their excitation energies. From the
strength fuction, we can determine its moments. Here we are interested in the
zeroth, first and second order moments\cite{Kur96,Boh96}:\[
S_{0}^{(i)}=\sum _{n\neq 0}\left| \left\langle n\left| O^{(i)}\right| 0\right\rangle \right| ^{2}=\left\langle 0\left| O^{(i)\, 2}\right| 0\right\rangle -\left\langle 0\left| O^{(i)}\right| 0\right\rangle ,^{2}\]

\[
S_{1}^{(i)}=\sum _{n\neq 0}\left| \left\langle n\left| O^{(i)}\right| 0\right\rangle \right| ^{2}\varepsilon _{n}=\left\langle 0\left| O^{(i)}(H-E_{0})O^{(i)}\right| 0\right\rangle =\frac{1}{2}\left\langle 0\left| \left[ O^{(i)},\, \left[ H,O^{(i)}\right] \right] \right| 0\right\rangle ,\]
 \[
S_{2}^{(i)}=\sum _{n\neq 0}\left| \left\langle n\left| O^{(i)}\right| 0\right\rangle \right| ^{2}\varepsilon ^{2}_{n}=\left\langle 0\left| O^{(i)}(H-E_{0})^{2}O^{(i)}\right| 0\right\rangle =-\left\langle 0\left| \left[ H,O^{(i)}\right] \, \left[ H,O^{(i)}\right] \right| 0\right\rangle .\]
 Using the above moments, we can define the strength, mean energy and width
of the strength distribution of \( O^{(i)} \). The total strength is equal
to \( S_{0}^{(i)} \),\[
I^{(i)}=S_{0}^{(i)}.\]
 The mean energy is defined by\cite{Kur96,Boh96}\begin{equation}
\bar{\omega }^{(i)}=\frac{S_{1}^{(i)}}{S_{0}^{(i)}},
\end{equation}
 and the width is the square root of the variance of the strength distribution\[
\Gamma ^{(i)}=\sqrt{\sigma ^{(i)}}=\sqrt{\frac{S_{2}^{(i)}}{S_{0}^{(i)}}-\frac{S_{1}^{(i)\, 2}}{S_{0}^{(i)\, 2}}}.\]
 These quantities have a very simple physical interpretation in terms of the
doorway state associated with the excitation operator \( O^{(i)} \) defined
as

\begin{equation}
\left| \lambda _{i}\right\rangle  =  \frac{\left( O^{(i)}-\left\langle 0\left| O^{(i)}\right| 0\right\rangle \right) \left| 0\right\rangle }{\left\langle 0\right| \left( O^{(i)}-\left\langle 0\left| O^{(i)}\right| 0\right\rangle \right) ^{2}\left| 0\right\rangle ^{1/2}}
  =  \frac{\left( O^{(i)}-\left\langle 0\left| O^{(i)}\right| 0\right\rangle \right) \left| 0\right\rangle }{S_{0}^{(i)\, 1/2}}.
\end{equation}
 This state is orthogonal to the ground state,\[
\left\langle \lambda _{i}|0\right\rangle =0,\]
 and the zeroth and first order moments are written in terms of \( \left| \lambda _{i}\right\rangle  \)
as\begin{equation}
\label{str0}
S_{0}^{(i)}=\left| \left\langle \lambda _{i}\left| O^{(i)}\right| 0\right\rangle \right| ^{2}
\end{equation}
 and\begin{equation}
\label{str1}
S_{1}^{(i)}=\left| \left\langle \lambda _{i}\left| O^{(i)}\right| 0\right\rangle \right| ^{2}\bar{\omega }_{i,}
\end{equation}
 where the mean energy, $\bar\omega_i$, is the expectation value of \( (H-E_{0}) \) in the doorway
state,\begin{equation}
\label{meanE}
\bar{\omega }_{i}=\left\langle \lambda _{i}\left| (H-E_{0})\right| \lambda _{i}\right\rangle .
\end{equation}
 The equations (\ref{str0}) and (\ref{str1}) lead to the interpretation that
the doorway state exhausts the strength of \( O^{(i)} \). The width also has
a very simple physical interpretation: it is the square root of the dispersion
of the doorway state,\begin{equation}
\label{gamdisp}
\Gamma ^{(i)\, 2}=\sigma ^{(i)}=\left\langle \lambda _{i}\left| (H-E_{0})^{2}\right| \lambda _{i}\right\rangle -\left\langle \lambda _{i}\left| (H-E_{0})\right| \lambda _{i}\right\rangle ^{2}.
\end{equation}
 From equation (\ref{gamdisp}), we find that \[
\sigma ^{(i)}=\sum _{m}\left| \left\langle \lambda _{i}\left| H\right| m^{(i)}\right\rangle \right| ^{2},\]
 where the states \( \left| m^{(i)}\right\rangle  \) are orthogonal to the
ground and doorway states,\[
\left\langle m^{(i)}|\lambda _{i}\right\rangle =\left\langle m^{(i)}|0\right\rangle =0\qquad \mbox {for all}\; m^{(i)},\]
 Thus, in this definition, the square of the width is the transition probability
of the doorway state to the states in the orthogonal complement to the subspace
defined by the ground and the doorway states.

Now we specialize our discussion to the description of multiple giant dipole
resonances. In the case of a single giant dipole resonance (GDR), the excitation
operator is the dipole operator,\begin{equation}
\label{dipolop}
O^{(i)}=D=\frac{Z}{A}\sum _{i=1}^{N}z_{n}(i)-\frac{N}{A}\sum _{i=1}^{Z}z_{p}(i)=\frac{NZ}{A}\left( R_{zn}-R_{zp}\right) ,
\end{equation}
 where \( Z \), \( N \) and \( A \) denote the number of protons, neutrons
and their sum, respectively.

In this case, the moments and the doorway state are given by\begin{equation}
\label{s10}
S_{0}^{(1)}=\left\langle 0\left| D^{2}\right| 0\right\rangle ,
\end{equation}
\begin{equation}
\label{s11}
S_{1}^{(1)}=\frac{1}{2}\left\langle 0\left| \left[ D,\, \left[ H,D\right] \right] \right| 0\right\rangle ,
\end{equation}
\begin{equation}
\label{s12}
S_{2}^{(1)}=-\left\langle 0\left| \left[ H,D\right] \, \left[ H,D\right] \right| 0\right\rangle ,
\end{equation}
\begin{equation}
\label{lam1}
\left| \lambda _{1}\right\rangle =\frac{D\left| 0\right\rangle }{\left\langle 0\right| D^{2}\left| 0\right\rangle ^{1/2}},
\end{equation}
 since parity conservation and the property that \( D \) is an odd-parity operator
imply that \( \left\langle 0\right| D\left| 0\right\rangle =0 \).

As will become clear in the development of the paper, the excitation operator
for the double giant dipole resonance (DGDR) should be a linear combination
of \( D^{2} \) and \( D \),\[
O^{(2)}=D^{2}+C_{1}^{(2)}D,\]
 such that the doorway states for DGDR and GDR are orthogonal,\[
\left\langle \lambda _{2}|\lambda _{1}\right\rangle =0.\]
 In this particular case, parity leads to \( C_{1}^{(2)}=0 \) and the excitation
operator for the DGDR is the usual one \cite{Kur96}, \( O^{(2)}=D^{2} \).
Therefore, the moments and the doorway state are\begin{equation}
\label{s20}
S_{0}^{(2)}=\left\langle 0\left| D^{4}\right| 0\right\rangle -\left\langle 0\left| D^{2}\right| 0\right\rangle ^{2},
\end{equation}
\begin{equation}
\label{s21}
S_{1}^{(2)}=\frac{1}{2}\left\langle 0\left| \left[ D^{2},\, \left[ H,D^{2}\right] \right] \right| 0\right\rangle ,
\end{equation}
\begin{equation}
\label{s22}
S_{2}^{(2)}=-\left\langle 0\left| \left[ H,D^{2}\right] \, \left[ H,D^{2}\right] \right| 0\right\rangle 
\end{equation}
\begin{equation}
\label{lam2}
\left| \lambda _{2}\right\rangle =\frac{\left( D^{2}-\left\langle 0\left| D^{2}\right| 0\right\rangle \right) \left| 0\right\rangle }{\left\langle 0\right| \left( D^{2}-\left\langle 0\left| D^{2}\right| 0\right\rangle \right) ^{2}\left| 0\right\rangle ^{1/2}}.
\end{equation}

Now we come to the triple giant dipole resonance (TGDR). By the same criterion,
the excitation operator should be a linear combination of the form \[
O^{(3)}=D^{3}+C_{2}^{(3)}D^{2}+C_{1}^{(3)}D,\]
 such that the doorways are orthogonal:\[
\left\langle \lambda _{3}|\lambda _{1}\right\rangle =\left\langle \lambda _{3}|\lambda _{2}\right\rangle =0.\]
 These orthogonality conditions together with parity give\[
C_{2}^{(3)}=0\qquad \mbox {and}\qquad C_{1}^{(3)}=-\frac{\left\langle 0\left| D^{4}\right| 0\right\rangle }{\left\langle 0\left| D^{2}\right| 0\right\rangle }.\]
 Therefore, the excitation operator for the triple giant dipole resonance (TGDR)
is\[
O^{(3)}=D^{3}-D\frac{\left\langle 0\left| D^{4}\right| 0\right\rangle }{\left\langle 0\left| D^{2}\right| 0\right\rangle }.\]
 As in the previous case, we can write the moments and the doorway state of
the TGDR excitation operator \( O^{(3)} \) in terms of powers of \( D \).
The expressions are cumbersome and are written below,\begin{equation}
\label{s30}
S_{0}^{(3)}=\left\langle 0\left| D^{6}\right| 0\right\rangle -\frac{\left\langle 0\left| D^{4}\right| 0\right\rangle ^{2}}{\left\langle 0\left| D^{2}\right| 0\right\rangle },
\end{equation}
\begin{eqnarray}
S_{1}^{(3)} & =\frac{1}{2} & \left\{ \left\langle 0\left| \left[ D^{3},\, \left[ H,D^{3}\right] \right] \right| 0\right\rangle \right. \label{s31} \\
 & - & \frac{\left\langle 0\left| D^{4}\right| 0\right\rangle }{\left\langle 0\left| D^{2}\right| 0\right\rangle }\left( \left\langle 0\left| \left[ D,\, \left[ H,D^{3}\right] \right] \right| 0\right\rangle +\left\langle 0\left| \left[ D^{3},\, \left[ H,D\right] \right] \right| 0\right\rangle \right) \nonumber \\
 & + & \left. \frac{\left\langle 0\left| D^{4}\right| 0\right\rangle ^{2}}{\left\langle 0\left| D^{2}\right| 0\right\rangle ^{2}}\left\langle 0\left| \left[ D,\, \left[ H,D\right] \right] \right| 0\right\rangle \right\} ,\nonumber 
\end{eqnarray}
\begin{eqnarray}
S_{2}^{(3)} & =- & \left\{ \left\langle 0\left| \left[ H,D^{3}\right] \, \left[ H,D^{3}\right] \right| 0\right\rangle \right. \label{s32} \\
 & - & \frac{\left\langle 0\left| D^{4}\right| 0\right\rangle }{\left\langle 0\left| D^{2}\right| 0\right\rangle }\left( \left\langle 0\left| \left[ H,D\right] \, \left[ H,D^{3}\right] \right| 0\right\rangle +\left\langle 0\left| \left[ H,D^{3}\right] \, \left[ H,D\right] \right| 0\right\rangle \right) \nonumber \\
 & + & \left. \frac{\left\langle 0\left| D^{4}\right| 0\right\rangle ^{2}}{\left\langle 0\left| D^{2}\right| 0\right\rangle ^{2}}\left\langle 0\left| \left[ H,D\right] \, \left[ H,D\right] \right| 0\right\rangle \right\} ,\nonumber 
\end{eqnarray}
\begin{equation}
\label{lam3}
\left| \lambda _{3}\right\rangle =\frac{\left( D^{3}-D\frac{\left\langle 0\left| D^{4}\right| 0\right\rangle }{\left\langle 0\left| D^{2}\right| 0\right\rangle }\right) \left| 0\right\rangle }{\left( \left\langle 0\left| D^{6}\right| 0\right\rangle -\frac{\left\langle 0\left| D^{4}\right| 0\right\rangle ^{2}}{\left\langle 0\left| D^{2}\right| 0\right\rangle }\right) ^{1/2}}.
\end{equation}
 Our scheme leads to the usual choice for the single and double giant dipole
resonances but to a different one for the triple giant resonance. Had we adopted
the naive choice,\begin{equation}
\label{dumD3}
\tilde{O}^{(3)}=D^{3},
\end{equation}
 the moments would be given by the first terms of Eqs. (\ref{s30}), (\ref{s31}),
and (\ref{s32}) and the doorway state would have been\[
\left| \tilde{\lambda }_{3}\right\rangle =\frac{D^{3}\left| 0\right\rangle }{\left\langle 0\right| D^{6}\left| 0\right\rangle ^{1/2}}.\]

Up to this point, our discussion has been model independent. In the following,
we will use the Tomonaga model\cite{Tom55}, to show that our choice for the
TGDR excitation operator leads to the expected results. On the other hand, we
will see that we would obtain unphysical results if we had adopted the naive
choice for the TGDR excitation operator.

A brief introduction to the Tomonaga model is given in Ref. \cite{Kur96}. Here,
we give a brief presentation of it to expose the key points which make possible
to extract very simple relationships between strengths, mean energies and widths
of the multiple giant dipole resonances.

First we observe that, given the dipole operator of Eq. (\ref{dipolop}), we
find the momentum conjugate to it to be\cite{Kur96}\[
P=\frac{A}{NZ}\left( \frac{Z}{A}\sum _{i=1}^{N}p_{zn}(i)-\frac{N}{A}\sum _{i=1}^{Z}p_{zp}(i)\right) =\frac{A}{NZ}\left( \frac{Z}{A}P_{zn}-\frac{N}{A}P_{zp}\right) ,\]
 which satisfy\[
\left[ D,P\right] =i\hbar .\]
 The next step in the construction of the model is to perform a canonical transformation
to \( D \), \( P \), and intrinsic variables and expand the Hamiltonian up
to terms quadratic in \( D \) and \( P \). This expansion can be written in
a very useful way if we introduce the operators that create and destroy dipole
bosons as\[
D=\frac{b}{\sqrt{2}}\left( c_{d}^{\dagger }+c_{d}\right) \qquad \mbox {and}\qquad P=\frac{i\hbar }{b\sqrt{2}}\left( c_{d}^{\dagger }-c_{d}\right) ,\]
 where \( b \) is the characteristic length of the dipole excitation.

Thus, the Hamiltonian can be expanded as \begin{eqnarray}
H & = & H_{00}+H_{10}\, c_{d}^{\dagger }+H_{01}\, c_{d}\\
 & + & \frac{1}{2}\left( H_{20}\, c_{d}^{\dagger \, 2}+2H_{11}\, c_{d}^{\dagger }c_{d}+H_{02}\, c_{d}^{2}\right) \label{TomH} 
\end{eqnarray}
 where the \( H_{ij}=H_{ji}^{\dagger} \) depend only on the intrinsic variables.

In order for the ground state in the Tomonaga theory to be the direct product
of the vacuum of the dipole boson and an intrinsic ground state\cite{Kur96},
\( \left| 0\right\rangle =\left| 0\right\rangle _{d}\otimes \left|
  0\right\rangle _{I}\), we impose the condition that
\( H_{10}\left| 0\right\rangle =H_{20}\left| 0\right\rangle =0\), with
\(c_d\left|0\right\rangle =0\).
 Given our definitions, the doorway states for the single, double and triple
giant dipole resonances are, respectively, the one, two and three phonon states,\begin{eqnarray*}
\left| \lambda _{1}\right\rangle  & = & c_{d}^{\dagger }\left| 0\right\rangle ,\\
\left| \lambda _{2}\right\rangle  & = & \frac{c_{d}^{\dagger \, 2}}{\sqrt{2!}}\left| 0\right\rangle ,\\
\left| \lambda _{3}\right\rangle  & = & \frac{c_{d}^{\dagger \, 3}}{\sqrt{3!}}\left| 0\right\rangle .
\end{eqnarray*}
 The total strengths are \[
S_{0}^{(1)}=\frac{1}{2}b^{2},\qquad \qquad S_{0}^{(2)}=\frac{1}{2}b^{4},\qquad \mbox {and}\qquad S_{0}^{(3)}=\frac{3}{4}b^{6},\]
 which lead to the relations\[
S_{0}^{(2)}=2\, S_{0}^{(1)\, 2}\qquad \mbox {and}\qquad S_{0}^{(3)}=6\, S_{0}^{(1)\, 3}.\]
 The mean energies calculated according to Eqs. (\ref{meanE}) and (\ref{TomH})
are\[
\bar{\omega }_{n}=n\, \left\langle 0\left| H_{11}\right| 0\right\rangle ,\]
 with \( n=1,2,3 \) for the single, double, and triple GDR,respectively, which
immediately implies that they satisfy the harmonic limit,\[
\bar{\omega }_{2}=2\bar{\omega }_{1}\qquad \mbox {and}\qquad \bar{\omega }_{3}=3\bar{\omega }_{1}.\]

Had we adopted the naive form of the triple giant dipole excitation operator,
given in Eq. (\ref{dumD3}), we would have found\[
\left| \tilde{\lambda }_{3}\right\rangle =\sqrt{\frac{2}{5}}\left| \lambda _{3}\right\rangle +\sqrt{\frac{3}{5}}\left| \lambda _{1}\right\rangle .\]
 That is, the doorway state would have been a linear combination of the three-
and one-phonons states. The total strength and mean energy of the TGDR would
then be \[
\tilde{S}_{0}^{(3)}=\frac{15}{8}b^{6}=15S_{0}^{(1)\, 3},\]
 and\[
\tilde{\omega }_{3}=\frac{9}{5}\left\langle 0\left| H_{11}\right| 0\right\rangle =\frac{9}{5}\bar{\omega }_{1}.\]
 The naive choice of the TGDR excitation operator thus leads to a mean excitation
energy smaller than that obtained for the double GDR, in complete disagreement
with the expected result. Of course, this result is a consequence of the fact
that, with the naive choice, the doorway state is a linear combination of one
and three phonon states.

To calculate the widths of the giant dipole excitations, we use Eq. (\ref{gamdisp}),
which expresses the widths in terms of the dispersion of the doorway states.
We find\begin{eqnarray*}
\sigma ^{(n)} & = & n\, \left\langle 0\left| H_{10}H_{01}\right| 0\right\rangle +\frac{1}{4}n(n-1)\left\langle 0\left| H_{20}H_{02}\right| 0\right\rangle \\
 & + & n^{2}\left( \left\langle 0\left| H_{11}^{2}\right| 0\right\rangle -\left\langle 0\left| H_{11}\right| 0\right\rangle ^{2}\right) ,
\end{eqnarray*}
 with \( n=1,2,3 \) for the single, double and triple GDR, respectively. If
we neglect the quadratic terms \( H_{20} \) and \( H_{02} \), and the dispersion
in \( H_{11} \), we find that \[
\Gamma ^{(2)}=\sqrt{2}\Gamma ^{(1)}\qquad \mbox {and}\qquad \Gamma ^{(3)}=\sqrt{3}\Gamma ^{(1)}.\]
 In general, however, we have\[
\Gamma ^{(2)}>\sqrt{2}\Gamma ^{(1)}\qquad \mbox {and}\qquad \Gamma ^{(3)}>\sqrt{3}\Gamma ^{(1)}.\]

Following Ref. \cite{Kur96}, we can discuss in a model independent way the
consequences of the hypotheses that charge exchange and quadratic momentum-dependent
interactions are absent.

These two hypotheses imply that \( [D,[V,D]]=0 \) and, since \( [D,[T,D]] \)
is a constant equal to\[
[D,[T,D]]=\frac{\hbar ^{2}}{m}\frac{NZ}{A},\]
 one has that \( S_{1}^{(1)} \) is equal to the Thomas-Reiche-Kuhn value\cite{Rin80},\[
S_{1}^{(1)}=\frac{\hbar ^{2}}{2m}\frac{NZ}{A}=S_{TRK}^{(1)}.\]

Assuming that \( [D,[V,D]]=0 \) as we did in the case of the GDR, we easily
calculate the commutators that appear in the expressions for \( S_{1}^{(2)} \)
and \( S_{1}^{(3)} \), Eqs. (\ref{s21}) and (\ref{s31}), to find\[
S_{1}^{(2)}=4\left\langle 0\left| D^{2}\right| 0\right\rangle S_{TRK}^{(1)},\]
 \[
S_{1}^{(3)}=\left( 3\left\langle 0\left| D^{4}\right| 0\right\rangle +\frac{\left\langle 0\left| D^{4}\right| 0\right\rangle ^{2}}{\left\langle 0\left| D^{2}\right| 0\right\rangle ^{2}}\right) S_{TRK}^{(1)}.\]
 The expectation values of \( D^{2} \) and \( D^{4} \) in the ground state
can be easily expressed in terms of the single and double GDR strengths, using
Eqs. (\ref{s10}) and (\ref{s20}) ,to write\[
S_{1}^{(2)}=4S_{0}^{(1)}S_{TRK}^{(1)},\]
 \[
S_{1}^{(3)}=\frac{\left( S_{0}^{(2)}+S_{0}^{(1)\, 2}\right) \left( S_{0}^{(2)}+4S_{0}^{(1)\, 2}\right) }{S_{0}^{(1)\, 2}}S_{TRK}^{(1)}.\]
 From these equations, we find model independent relations between the mean
energies and strengths,\[
\frac{\bar{\omega }_{2}}{\bar{\omega }_{1}}=\frac{4S_{0}^{(1)\, 2}}{S_{0}^{(2)}},\]
 \[
\frac{\bar{\omega }_{3}}{\bar{\omega }_{1}}=\frac{\left( S_{0}^{(2)}+S_{0}^{(1)\, 2}\right) \left( S_{0}^{(2)}+4S_{0}^{(1)\, 2}\right) }{S_{0}^{(3)}S_{0}^{(1)}}.\]

If we make the additional hypothesis that the mean energies satisfy the harmonic
limit, we find that the strengths obey the relations, \( S_{0}^{(2)}=2S_{0}^{(1)\,2} \),
\( S_{0}^{(3)}=6S_{0}^{(1)\,3} \), consistent with the Tomonaga model. On the
other hand, had we adopted the naive choice \( O^{(3)}=D^{3}, \) the latter
of these relations would be modified to\[
\frac{\tilde{\omega }_{3}}{\bar{\omega }_{1}}=9\frac{\left( S_{0}^{(2)}+S_{0}^{(1)\, 2}\right) S_{0}^{(1)}}{\tilde{S}_{0}^{(3)}},\]
 and in the harmonic limit we would have \( \tilde{S}_{0}^{(3)}=9S_{0}^{(1)\, 3} \),
inconsistent with the relation \( S_{0}^{(3)}=6S_{0}^{(1)\, 3} \).

To summarize, we have shown that when using sum rules to probe the properties
of the TGDR, the excitation operator must be modified from the naive choice
\( D^{3}. \) In the Tomonaga model, our choice for the TGDR excitation
operator gives values for the strength and mean energy in accordance with the
harmonic limit and a width greater than \( \sqrt{3} \) times that of the GDR.
On the other hand, had we adopted the naive choice \( D^{3} \), we would obtain
unphysical results, namely, that the strength is fifteen times the GDR strength
cubed and the mean energy is nine fifths the GDR energy. 

Assuming the absence of a charge exchange component in the nuclear interaction,
we derive model independent relationships between the mean energy and strengths
of the multiple giant resonances, which extend the relationships derived in
Ref. \cite{Kur96} for the DGDR.

Our approach is very general and can be applied to multiple giant resonances
of any multipolarity. For multiple giant quadrupole resonances, for example,
we would find a modification of the excitation operator already at the DGQR
level. Here we do not extend the analysis to the quadrupole resonance, since
it is beyond the scope of the present paper.

\end{document}